# Heat shock proteins may be a missing link between febrile infection and cancer tumor rejection via autoantigen molecular mimicry

Amin Zia, PhD

dYcode Research

amin.zia@dYcode.bio

Toronto, Ontario, Canada L6C 2R9

## ABSTRACT

Numerous epidemiological studies suggest febrile infections could confer long-term immunity to certain types of cancers, though the precise mechanisms for this phenomenon remain unclear. Systemic heat-shock responses to fever may be key to understanding the overlapping outcomes of immune responses to infection and cancer. To investigate this hypothesis, we performed epitope discovery between heat-shock proteins (HSP) and cancer-associated antigens (CAA) and annotated the results with experimentally validated epitopes in the Immune Epitope Database (IEDB) (Vita et al., 2019). Further, epitopes were matched with their homologs in human pathogens. Results identified 94 epitopes shared between HSPs and CAAs, with experimental evidence of presentation at MHC molecules and with high homology to several epitopes of human pathogens. The identified epitopes can be used as candidates for designing cancer vaccines. They may also be used to identify autoreactive antibodies or TCR specificities that, as antibody drugs and cell therapies, would reproduce the effect of febrile infection in conferring cancer immunity. Our results support the hypothesis that the loss of self-tolerance to HSPs during febrile infection confers tumor immunity through molecular mimicry.

## INTRODUCTION

*Infection and Cancer:* The relationship between infection and cancer is complex. Chronic infections could initiate growth of cancer cells (e.g., Human papillomavirus (HPV) and cervical cancer, Heliobacteria pylori and gastric cancer, and Epstein-Bars virus (EBV) and B-cell lymphoma) and are estimated to cause 15%-21% of all cancers (Çuburu et al., 2022; zur Hausen & de Villiers, 2014). Conversely, oncolytic viruses selectively target tumor cells and have been examined as cancer therapeutics in several clinical trials (Shalhout et al., 2023). Additionally, pathogens in the human microbiome prove essential for maintaining homeostasis and are associated with both prevention and initiation of cancers (Villemin et al., 2023). The present study examines an alternative relationship that associates febrile infections with cancer protection and even, in rare cases, spontaneous remission (SR) of tumors (Hobohm, 2005; Radha & Lopus, 2021).

*Coley's Toxins:* In the late 18th century, a young surgeon at the New York Cancer Hospital named Dr. William Coley was intrigued by numerous case reports documenting episodes of spontaneous tumor remission in cancer patients who experienced severe febrile infections (Coley, 1891). Motivated by a personal tragedy, Coley investigated the clinical feasibility of treating cancer by inducing infection in patients using a mixture of bacterial toxins (e.g. Streptococcus pyogenes and Serratia marcescens), later known as Coley's mixture (McCarthy, 2006). In the following decades and using this mixture, Coley treated hundreds of patients with soft-tissue sarcomas, lymphomas, ovarian carcinoma, cervical carcinoma, testicular and renal tumors and breast carcinoma (Coley, 1910; Kienle, 2012; Starnes, 1992). These treatments achieved unprecedented success when no other treatment alternatives existed. According to one report, 51.9% of Coley's patients with inoperable soft-tissue sarcomas had complete tumor regression and survived for more than five years, of which 21.2% had no clinical evidence of tumor for at least 20 years (Starnes, 1992). A recent prospective study reported Coley's method had superior remission results in terms of overall survival compared to modern chemotherapy and radiotherapy in kidney, ovarian, and breast cancers and soft-tissue sarcomas (Kienle, 2012). Despite this success, with the birth of new cancer treatments, including chemotherapy and radiotherapy, and due to anectodical mortalities caused by infection toxicities (Wei et al., 2008), Coley's method never reached FDA approval and was practically forgotten by the scientific community. Nevertheless, Coley's work became a foundation for our understanding of the interplay between immune responses to infection and cancer and served as first examples of modern cancer immunotherapy (*Footnote: Coley's treatment was used routinely until 1963 for the treatment of sarcomas* (Richardson et al., 1999). *The last recorded successful application of the toxin was in China in 1980, where symptoms completely disappeared in a terminal liver cancer patient* (Jessy, 2011)).

*Spontaneous Remission:* Although extremely rare (with an estimate of 1 in 60,000-100,000 in all cancers (Black et al., 1956; Boyd, 1957; Dobosz & Dzieciątkowski, 2019) and 1 in 400 in melanoma (Botseas, 2012)), spontaneous remission (SR) of tumors is a well-documented phenomenon reported in numerous cases of several types of cancers, including lung, lymphoma, leukemia, retinoblastoma, neuroblastoma, melanoma, choriocarcinoma, renal, bladder, gastrointestinal, sarcoma and carcinoma of the urinary bladder, renal adenocarcinoma, breast cancer and lung carcinoma (Radha & Lopus, 2021). Although the exact mechanism of SR is unclear, febrile infections have been proposed as the initiating factor in approximately 9% of these case reports (Minacapelli et al., 2022), where partial or complete tumor regression is attributed to reactivation of an immune response to acute infection during standard treatment. More recently, during the Covid-19 pandemic, several groups reported spontaneous remission of tumors due to SARS-CoV-2 infection (Meo et al., 2023), including in Hodgkin lymphoma (Challenor & Tucker, 2021), chronic lymphocytic leukemia (CLL) (Bülbül et al., 2022), and NK lymphoma (Pasin et al., 2020).

*Long-term cancer immunity due to infections:* Beyond radical remission of tumors in spontaneous remission, overwhelming evidence from epidemiological studies suggests acute infections may confer long-term cancer protection through cancer immunosurveillance, highlighting the potential involvement of adaptive immunity. A 2013 study of 700 chronic lymphoid leukemia patients in Italy showed one infection in childhood (measles, chickenpox, rubella, mumps, pertussis) significantly reduces the chance of leukemia in adulthood (with an odd ratio (OR) of 0.66), and multiple infections reduce this chance even further (Parodi et al., 2013), though interestingly, not for multiple myeloma (Stagnaro et al., 2018). A follow-up study of approximately 1100 Italian individuals found an inverse association of B-cell lymphomas with past infections of rubella (OR 0.80), pertussis (OR 0.74), and any infection (OR 0.75), with a negative trend by number of infections (OR 0.66 for three infections or more) (Parodi et al., 2020). A 2003 study of 600 melanoma patients in 6 European countries showed children with one or more severe infections (with fever > 38.5°C, e.g. sepsis, pneumonia, pulmonary tuberculosis and Staphylococcus aureus infection) had a reduced chance of developing melanoma in adulthood (OR of 37%) (Krone et al., 2003). A 2010 study of 360 women showed childhood mumps parotitis reduced the chance of developing ovarian cancer in adulthood by an OR of 80% (Cramer et al., 2010a). A 20-year study of 1,000 children in the UK found attending daycare and repeated exposure to respiratory infections reduced the risk of developing lymphoma in adulthood to OR of 48% (Gilham et al., 2005). In a 2006 review study, childhood infections were shown to correlate with a reduced risk of developing multiple cancers, including melanoma (Hoption Cann et al., 2006). Interestingly, a recent research demonstrates personal history of acute infections may also have a significant contribution to the efficacy of immunotherapy, e.g., through modification of the expression of therapeutic targets, preexisting immunity to therapies such as EBV-specific CAR T cells, or shared antigens (for a review see (Jacqueline et al., 2018)).

*Cancer immunity due to infectious disease vaccines:* In addition to the history of infection, a history of vaccination against infectious diseases is correlated with protection against certain cancers. In 1976, Morales, Eidinger and Bruce (Morales et al., 2017) observed episodes of spontaneous remission of bladder cancer tumors in patients after inoculation of the tuberculosis vaccine (Mycobacterium bovis), which was later called Bacillus Calmette-Guérin (BCG) vaccine (Mukherjee et al., 2022). Since then, many large-scale long-term studies associated reduced chances of cancer development to previous vaccinations, including reduced risk of lymphoma and leukemia after BCG and smallpox vaccination (Marron et al., 2021; Villumsen et al., 2009), reduced risk of lung cancer after BCG vaccination (Usher et al., 2019) and reduced risk of acute lymphoblastic leukemia after Haemophilus influenzae type-B vaccination (Marron et al., 2021). The therapeutic BCG vaccine, injected directly into the bladder, was later approved by the FDA and is currently the standard of care in adjuvant therapy for treating high-risk patients with superficial non-muscular urinary bladder cancer (Jiang & Redelman-Sidi, 2022). Subsequent clinical studies showed the efficacy of this vaccine for treating other cancers, including acute lymphoblastic leukemia (Mathé et al., 1969) and melanoma (Leong, 1996). Several other infectious disease vaccines have been repurposed and are in clinical trials as vaccines for a wide range of cancers (Vandeborne et al., 2021).

*Innate Immunity:* In the context of an established cancer, innate immune responses to infection comprise several components which may contribute to tumor rejection. Pathogen-associated molecular patterns (PAMPs) of damaged or dying cells due to infection activate danger signals, releasing damage-associated molecular patterns (DAMPs) and initiating a cascade of anti-tumor interferon and proinflammatory cytokine signaling by lymphoid and myeloid lineage cells (Janeway & Medzhitov, 2002). Similarly, the excess accumulation of cytosolic double-stranded DNA (dsDNA) (Li & Chen, 2018) and double-stranded RNA (dsRNA) (Liu et al., 2015) originating from accelerated apoptosis and phagocytosis and the ectopically expressed endogenous retroviral particles (ERVs) in tumor microenvironment, may further activate DAMP signals (Lee et al., 2020). The cascade of these events could prime an innate immune response in the tumor microenvironment (Jacqueline, Parvy, et al., 2020). These ideas are at the core of current treatments which pose to prime an innate immune response by stimulants such as mistletoe extract (Reuter et al., 2018) or PAMP/DAMPs as vaccine adjuvants (Sun et al., 2021).

*Adaptive Immunity:* The long-term cancer protection due to past infections may, however, be explained more accurately in the context of an adaptive immunity. From this perspective, infection and cancer involve different signaling cascades. In infections, exogenous pathogens commonly induce type-II responses; in cancer, antigens are typically endogenous and activate a type-I response. Nevertheless, their protective outcomes may overlap (e.g. through cross-presentation pathways). While the precise mechanism for this proposition remains unclear, several studies suggest molecular mimicry a key to this functional overlap.

*Molecular Mimicry* is defined as the pathogenesis of disease due to the similarity of epitopes of human proteins and pathogens. The term molecular mimicry was first used in the context of rheumatic fever when antibody titers to group-A streptococcal virus were discovered to have cross-reactivity to human heart tissues (Kaplan & Meyersian, 1962). Screening of 600 monoclonal antibodies raised against 11 different pathogens found 3.5% of them reacted to self-antigens in normal uninfected mice (Srinivasappa et al., 1986). Since then, overwhelming evidence has accumulated, demonstrating involvement of molecular mimicry in infections in the pathogenesis of several autoimmune disorders, such as in multiple sclerosis (MS), type-1 diabetes (T1D), ankylosing spondylitis, myasthenia gravis, systemic lupus erythematosus, autoimmune myocarditis and many others (Cusick et al., 2012; W. Zhang & Reichlin, 2008); for review see (Johnson & Jiang, 2022), and for one of the most interesting recent examples, refer to (Lanz et al., 2022)].

*Molecular Mimicry in the context of cancer* resulting in spontaneous tumor remission has been proposed, though never experimentally demonstrated (Sioud, 2002). Beyond spontaneous remission, the first experimental evidence of molecular mimicry in cancer was reported (Loftus et al., 1996), in which several peptide epitopes from viral pathogens (the Herpes simplex virus-1, the Herpes simplex virus-2, the Pseudorabies virus, and the Adenovirus 3,7) were identified to have high sequence homology to melanoma antigen MART-1. These peptides sensitized the melanoma tumor cells by MART-1 specific cytotoxic T lymphocytes (CTL). In another study (Rubio-Godoy et al., 2002), a set of peptides were identified to be cross-reactive to the CTLs specific to melanoma antigen Melan-A. They showed a majority of these peptides originated from viral genomes and retained high sequence homology to self-peptides. Interestingly, the researchers highlighted previous results (Ogg et al., 1998) showing higher frequencies of Melan-A specific CTLs in cancer patients with autoimmune vitiligo (a disease of the destruction of skin melanocytes) and proposed CTL-reactive peptides with pathogen origin may contribute, through molecular mimicry, to the spontaneous appearance of vitiligo in metastatic melanoma patients (Nordlund et al., 1983). A recent study (Chiou et al., 2021) demonstrated evidence of viral mimicry between T cell epitopes of lung cancer tumor antigen TMEM161A and latent membrane protein 2a antigen of the Epstein-Barr virus (EBV), suggesting cross-reactivity of EBV-specific T cells may play a crucial role in controlling cancer progression and overcoming tumor antigen tolerance. In another study (Vujanovic et al., 2007), a Mycoplasma-derived peptide elicited CD4+ T cell responses against melanoma tumor antigen MAGE-A6, suggesting CD4+ T cell receptor (TCR) of antigen-specific CTLs cross-react to Mycoplasma. In a recent study (Manolio et al., 2022) authors identified epitopes of tumor-associated antigens which were highly homologous to HIV-derived epitopes, including in breast, prostate and colon cancers. They concluded memory CD8+ T cells elicited during the HIV infection might play a key role in controlling the development and progression of tumors, providing answers to long-standing observations HIV/AIDS patients have reduced incidence of these three cancers. An additional study (Choi et al., 2021) showed expression of the Epstein–Barr virus (EBV) signaling protein LMP1 in B cells provokes T cell responses to multiple melanoma tumor antigens and therefore induces protection through tumor surveillance by molecular mimicry (Iheagwara et al., 2014). See (Tagliamonte et al., 2023) for a review of other examples of molecular mimicry with bacterial epitopes in tumor associated antigens MART-1, MAGE-A6, MAGE-A10, MAGE-A3, MAGE-A1, MAGE-C2, MAGE-12, SSX-2 and MAGE-C1 and examples of molecular mimicry with viral epitopes of tumor antigens MART-1, TMEM161A, Gp100, CD274, HEPACAM, CEA, TELOMERASE, ISG15, MDK, C1QTNF12, CCR9, EPCAM, DPYSL2, CNIH4, and NDUFS2.

*Essential Role of Fever:* Sequence homology between epitopes in pathogens and self-proteins, as suggested by the above examples, does not entirely explain why these events are rare. Besides, experimental evidence suggests cross-reactive antibodies and T cells commonly exist in normal individuals without always breaking self-tolerance, and it is likely other complementing processes are required. Here, we hypothesize fever in acute infections acts as a central mediator for the molecular mimicry to become effective (Kleef et al., 2001). Indeed, according to Coley the induction of fever was the most crucial marker for spontaneous regression of cancer in his treatments. In a retrospective study of patients with inoperable sarcoma treated with Coley’s toxin, patients who experienced a fever (38-40°C) had an over 5-year survival rate higher than those who had no fever during treatment (Nauts, 1974). In addition, the majority of infections examined in epidemiological studies which correlate cancer immunity to infection are accompanied by fever (Kienle, 2012).

*Fever*, defined as an elevation in core temperature (38.5–40°C), is a beneficial protective response to infection (Jeffrey D. Hasday & Singh, 2000) in endothermic higher animals and many ectothermic vertebrates, arthropods, and annelids. The elevated core temperature, as a result of infection or otherwise induced, acts as an adjuvant to accentuate adaptive immune response through promoting antigen peptide presentation (Garstka et al., 2015), reprogramming of cytokine secretion, activation of proinflammatory signaling (e.g. TNF-a) and heat shock response (J. D. Hasday et al., 2001). The orchestrated heat shock response, in turn, induces the upregulation of cytoprotective heat shock proteins (HSP) in host cell tissue (Wang et al., 2020) and peripheral blood (Sonna et al., 2010). Therefore, HSPs may be the key to conferring cancer immunity in febrile infections (*Footnote: fever can also directly kill pathogens, inhibit their growth and induce the expression of immunogenic HSPs in pathogens which primes an innate immune response and inflammation; see* (Jeffrey D. Hasday & Singh, 2000) *for a review*.)

*Heat-shock Proteins* are families of ubiquitous and highly conserved proteins which act as housekeeping proteins under physiological conditions to maintain proper folding of the nascent polypeptide chain and as chaperonins to mediate protein folding and transport within the cell. During heat shock response, HSPs preserve essential cellular homeostasis by protecting component, function and chemical balance of tissue from chemical/ physiological stress and cells from injury (Dubrez et al., 2020). HSPs are mainly considered intracellular chaperones. However, HSPs also contribute to cell survival in the extracellular environment; present in the circulation of healthy individuals and prevalent in several pathological conditions such as infection and cancer (Chatterjee & Burns, 2017; Jay et al., 2022). Excess extracellular HSPs are released from damaged or stressed cells and appear to act as local “danger DAMP signals” which activate stress response programs in surrounding cells (Giuliano et al., 2011). Notably, the increased expression of extracellular HSPs due to stress (e.g., febrile infection) enhances the antigen uptake, sorts and processes antigen-presentation machinery (Murshid et al., 2012) and activates dendritic cells (Mcnulty et al., 2013). For a summary of HSP families, important members, and their cellular location and function see: (Chatterjee & Burns, 2017).

*HSPs are recognized by the immune system* and present prominent antigens in a wide range of diseases (Zügel & Kaufmann, 1999). For instance, immunogenic epitopes of HSP70 that elicit a CD4+ T cell response were identified in individuals with juvenile dermatomyositis (JDM) (Koffeman et al., 2011). Anti-HSP autoantibodies have been implicated in: (i) autism; (ii) connective tissue autoimmune disorders (Vojdani et al., 2004); (iii) non-small cell lung cancer patients (Zhong et al., 2003) and; (iv) the serum titer of anti-TRAP1 which is proposed as a biomarker for diagnosing small cell lung cancer and tumor immunity (Tan et al., 2023), to name a few. Activation of T cells recognizing HSP60 was shown to protect against experimental arthritis (Anderton et al., 1995), and HSP70-specific T cells were shown to regulate inflammation by production of suppressive cytokine IL10 in (Tanaka et al., 1999), while immunodominant T cell epitopes of HSP65 in Kawasaki Disease were identified (Sireci et al., 2000).

For several examples of the role of HSPs in different diseases, see (De Jong et al., 2014; Turturici et al., 2011; Willem Van Eden et al., 2005).

*Molecular Mimicry Involving HSP-derived epitopes* has also been implicated in several autoimmune disorders (De Jong et al., 2014). As an example antibodies to Campylobacter jejuni chaperones generated during infection may cross-react with the human HSP epitopes and reduce their neuroprotective function in peripheral nerve tissue (Loshaj-Shala et al., 2015), an event which may contribute to the pathogenesis of the rare autoimmune disorder Guillain-Barré syndrome (GBS) (Moise et al., 2016; Van Herwijnen et al., 2012). Immune responses to Mycobacterial HSP70 result in self-reactivity to a member of the human HSP70 protein family human implicated in rheumatoid arthritis (Shoda et al., 2016). In the context of cancer, however, HSPs are mainly considered cancer-associated antigens where their overexpression is proposed to promote cancer progression.

Here, we present computational evidence suggesting an alternative role for HSPs. Febrile infections induce a heat shock response in which the upregulation of HSPs enhances tissue homeostasis and promotes dendritic cell antigen presentation, inflammatory cytokine secretion and response to foreign pathogens. The immune system recognizes dysregulation of HSPs and anti-HSP antibody titers are detected in patients with infection (Bolhassani & Agi, 2019). In addition, the HSP-derived peptides constitutively presented at HLA type-I and type-II loci in normal tissues are also upregulated in inflamed tissues, and HSP-specific T cells are detected in patients with infection. We show that HSPs carry epitopes highly similar to cancer associated antigens and may mediate protection during febrile infection through molecular mimicry. The immunogenicity of these epitopes could be attributed to their highly conserved sequence homology to their microbial paralogs, and therefore the upregulation of HSPs in infections may facilitate antigen presentation of peptides from homolog sequences.

As a proof of principle, we focus on cancer-associated antigens (CAA) which consist of nonmutated self-antigens in tumor cells, e.g., MUC1 (Vlad et al., 2004). At least in the case of MUC1, research shows CAAs are temporarily upregulated in inflamed tissues during infection (Cramer et al., 2010b; Iheagwara et al., 2014), further justifying their study in this context. Furthermore, immune responses against CAAs are detected in premalignant precursors of various cancers, and, although less frequently, in healthy individuals, and are correlated with reduced cancer risk (Pinheiro et al., 2010; Tabuchi et al., 2016; Vella et al., 2009). In this context, we propose that HSPs may induce a cross-reactive adaptive immune response against CAAs and therefore confer cancer immunity.

## RESULTS

We sought to investigate whether the upregulation of HSPs during febrile infection and consequent immune response to these proteins confer long-term immunity to certain cancers (Fig. 1A). As a proof of concept exercise (Fig. 1B), a set of 178 cancer-associated antigens, and 281 HSPs and heat-shock factors were selected (see Methods), and a comprehensive peptide sequence search was conducted to identify epitopes shared between CAAs and HSP proteins. The research resulted in 2,491,869 pairs of peptide epitopes with lengths >= 7 amino acids and minimum similarity >85%. Each epitope pair consisted of a peptide from an HSP and a peptide from a CAA.

The peptide epitopes were further mapped onto The Immune Epitope Database (IEDB) (Vita et al., 2019), and any epitope which failed to have an exact amino-acid sequence match (with the same length or as a sub-sequence) with at least one epitope in IEDB was removed from further analysis (along with its paired epitope). The resulting set consisted of epitopes shared between HSPs and CAAs with records in IEDB with a matching gene. The supporting IEDB records provided experimental evidence for the epitopes' antibody reactivity, presented on MHC loci, or mediate T cell reactivity, and therefore were potentially immunogenic. This included 94 epitope pairs that were shared between 61 HSPs and factors and 32 CAAs were identified. For each epitope, a search was conducted for similar epitopes in 46 human pathogens (designated by the NCBI Human Pathogen Detection database (see Methods)) to further collect evidence for their potential immunogenicity. Any pathogen hit match was filtered with <7 alignment length and <85% amino-acid similarity. The statistical significance of each pair was calculated as ($p_{val} \sim 10^{-5}$) by computation simulations (see Methods). See Table-1 for a brief overview of identified.

In the following sections, we review a few notable epitopes with a focus on the HSP90 family. This family comprises one of the most abundant proteins in the eukaryotic cell constituting up to 1-2% of the cellular protein under physiological conditions. It is upregulated by several-fold during the response to febrile infection (Bolhassani & Agi, 2019) and enhances cytosolic translocation of extracellular antigens for cross-presentation by dendritic cells (Imai et al., 2011). Two distinct HSP90 genes - HSP90A1 and HSP90AB1 - encode the protein's inducible and constitutively expressed isoforms. Epitope mapping with sera from patients with a range of infections was shown to cross-react with HSP90-derived epitopes, including candidiasis, invasive aspergillosis, allergic bronchopulmonary aspergillosis, aspergilloma, Enterococcus faecalis, Corynebacterium jeikeium endocarditis and Streptococcus oralis septicemia (Al-Dughaym et al., 1994). Also, cross-reactive antibodies to extracellular HSP90 have been detected in the sera of patients with cancer and autoimmune disorders (Jay et al., 2022). Tumor cells are addicted to HSP90 family to maintain survival and grow in the harsh and stressed tumor microenvironment. Therefore, inhibition of HSP90 has recently gained traction with several candidates currently in clinical trials (Albakova et al., 2021; Den & Lu, 2012). Pharmacological inhibition of HSP90 compromises stability of proteins in tumor cells and therefore accelerates degradation through the proteasome activity. It also enhances the effectiveness of cancer immunotherapy by upregulation of the interferon response, and by enhancing the T-cell-mediated killing mechanisms (Haggerty et al., 2014; Mbofung et al., 2017).

While HSP90 is currently considered as a cancer associated antigen, our results extend the role of HSP90 family as mediators of protective cancer immunity in febrile infections. We identified nine epitope pairs between members of the HSP90 family (HSP90AA1, HSP90AB1, HSP90B1, and HSP90AB3P) and 9 CAAs (MMP2, TEK, ACTN4, ACTN1, EFTUD2, MUC1, NECTIN4,

GPC3, UBR4). We also identified four epitope pairs between HSP90 co-chaperones CDC37, CDC37L1 and CAAs OGT, MUC5AC, EEF2, and GSN and four epitope pairs shared between HSP90 activator AHSA1 and CAAs GPNMB, MAGEC1, CDC27, GPNMB (See Table 1, Fig. 2).

**MUC1 and HSP90AA1, HSP90AB1, and HSP90AB3P**

The epitope “SLEDPST” on MUC1 protein is highly similar to the epitope “SLEDPQT” (with one amino-acid difference) on HSP90AA1, HSP90AB1 and HSP90AB3P (see Fig. 2 and Fig. 3A). This epitope is located at the extracellular sea urchin sperm protein, enterokinase and agrin (SEA) domain of MUC1, neighbor to the normally hyperglycosylated repeat region (Gao et al., 2020). The paired HSP-derived epitope SLEDPQT falls into C-terminal domain, where HSP90s are involved in client protein interaction and binding, dimerization, and co-chaperone binding (Hoter et al., 2018).

The MUC1 epitope SLEDPST was identified as a substring match of an epitope presented on the HLA class II MHC molecule (IEDB epitope-2071116). The HSP90 epitope “SLEDPQT” matches 37 IEDB records with experimental evidence of presentation on MHC class-I and class-II loci.

The HSP-derive epitope “SLEDPQT” had high homology with 1066 species of human pathogens including Salmonella enterica (785 hits), Streptococcus agalactiae (19 hits), Burkholderia cepacian complex (2 hits), Escherichia coli (19 hits), Klebsiella pneumoniae (165 hits), Enterobacter hormaechei (39 hits), Enterobacter cloacae (4), Pseudomonas putida (2 hits), Enterococcus faecium (25 hits), Stenotrophomonas maltophilia (5 hits), and Acinetobacter baumannii (1 hit).

MUC1 is a protein-coding gene encoding mucin short variant S1 membrane-bound glycoprotein normally expressed in apical epithelial cells of lungs, stomach, intestines, eyes and other organs. The extracellular domain of the MUC1 protein is highly glycosylated in normal tissues. The overexpression of MUC1 in epithelial cells with hypo-glycosylation in this domain has been shown to promote tumor growth and progression of cancer and as a positive feedback for its overexpression (Cascio et al., 2017). MUC1 is overexpressed in many epithelial adenocarcinomas and is involved in promoting cell proliferation, inhibiting apoptosis and enhancing migration and invasion (Hollingsworth & Swanson, 2004; Nath & Mukherjee, 2014). It is proposed as a cancer-associated antigen in many cancers (Finn & Beatty, 2016; Lau et al., 2004), including pancreatic intraepithelial neoplasia, Intraductal papillary mucinous neoplasms, Barrett’s esophagus, adenomatous polyps, monoclonal gammopathy of undetermined significance and asymptomatic multiple myeloma, bronchial preneoplasia, breast ductal carcinomas (Vegt et al., 2007), pancreatic cancer (Roy et al., 2011; Winter et al., 2012) and epithelial ovarian cancer (Deng et al., 2013).

**GPC3 and HSP90AB1**

The epitope “DKKVLKV” on GPC3 was found to be highly similar to the HSP90AB1-derived epitope “DKKVEKV” (Fig. 2 and Fig. 3B). This epitope is presented on MHC class-II molecules, matched with IEDB epitope 1305673. The paired HSP90AB1-derived epitope “DKKVEKV” was matched with 10 IEDB records that show the MHC class I and II presentation. Of particular interest, this epitope was matched with an IEDB epitope 27000 which indicates positive qualitative T-cell binding in the context of HLA-A2.

The HSP90AB1 epitope was shared with a total of 338 species from human pathogens, including Clostridioides difficile (231 hits), Salmonella enterica (12 hits), Enterobacter hormaechei (6 hits), Vibrio vulnificus (1 hit), Acinetobacter baumannii (1 hit), Legionella pneumophila (5 hits), Clostridium botulinum (6 hits), Escherichia coli (2 hits), Enterobacter kobei (2 hits), Pseudomonas putida (1 hit), Streptococcus pneumoniae (52 hits), Streptococcus agalactiae (4 hits), Flavobacterium psychrophilum (12 hits), Bacillus cereus group (1 hit).

Interestingly, HSP90B epitope IL***DKKVEKV*** was previously shown to be shared in melanoma cell lines (Jarmalavicius et al., 2012), with sequence homology with measles virus MHC class-I epitopes that induces auto-reactive CD8+ T cells. Furthermore, when incubated with viral epitopes homologous to this epitope, the melanoma-specific CD8+ T cells were expanded *in vitro* (Chiaro et al., 2021).

GPC3 (Glypican-3) is a member of the cell surface proteoglycan glypican family of glypican-related integral membrane proteoglycan family (GRIPS) which regulate various signaling pathways during development and tissue homeostasis. Glypican-3 is overexpressed in several cancers, including hepatocellular carcinoma (HCC), melanoma, and lung cancer and plays a role in tumor development and progression. GPC3 is a promising target for cancer therapy and several therapeutic agents targeting GPC3 for the treatment of various cancers are currently in clinical trials (J. Zhang et al., 2018).

**MMP2 and HSP90B1**

The epitope “YSLFLVAAH” found in MMP2 was similar to the HSP90B1-derived epitope “YSAFLVADK” (Fig. 2 and Fig. 3C) and presented on the MHC class-II locus (two IEDB records). The HSP90B1-derived epitope was also presented on the MHC Class-II locus (IEDB epitope 642641). This epitope was a homolog to 29 human pathogens, including Klebsiella pneumoniae (10 hits), Flavobacterium psychrophilum (1 hit), Enterococcus faecium (1 hit), Clostridium botulinum (2 hits), Serratia marcescens (1 hit), Escherichia coli (2 hits), Pseudomonas aeruginosa (4 hits), Shigella (2 hits), and Salmonella enterica six hits).

Interestingly, the MMP2 epitopes “RLSQDDI” and “FNGKEYN” were found highly similar to epitopes “RLSKDDI” from HSPA2 and “FNGKELN” from heat shock factors HSPA8, HSPA6, and HSPA2, respectively. These HSP epitopes had numerous hits in human pathogens, including Acinetobacter baumannii, Aeromonas veronii, Citrobacter freundii, Corynebacterium striatum, Enterobacter hormaechei, Enterococcus faecalis, Enterococcus faecium, Escherichia coli, Klebsiella pneumoniae, Listeria monocytogenes, Mycobacterium tuberculosis, Pseudomonas aeruginosa, Pseudomonas putida, Salmonella enterica, Shewanella algae, Staphylococcus aureus, Vibrio parahaemolyticus, and Vibrio vulnificus.

MMP2 (matrix metalloproteinase-2) is a member of the zinc-dependent endopeptidase family which plays a crucial role in the remodeling of the extracellular matrix (ECM). MMP2 is involved in various physiological and pathological processes such as embryonic development, angiogenesis, wound healing, and cancer progression. MMP2 is overexpressed and associated with tumor invasion, metastasis, and angiogenesis by degrading the ECM components, such as type IV collagen, a major component of the basement membrane. Therefore, MMP2 is considered a potential therapeutic target for cancer treatment (Egeblad & Werb, 2002)

**NECTIN4 and HSP90AA1, HSP90AA4P**

We found NECTIN4-derived epitope “SAAVTSE” as highly similar to the HSP90 epitope “SAAVTEE” (Fig. 2 and Fig. 3D). The NECTIN4 epitope matched the IEDB epitope 1282685 and presented at MHC Class II. The HSP90-derived epitope was matched with 7 IDEB epitopes presented on MHC class I and class II.

The HSP90 epitope “SAAVTEE” was highly similar to 59 human pathogen epitopes, including Pseudomonas aeruginosa (36 hits), Bacillus cereus group (19), Streptococcus agalactiae (3 hits), and Enterococcus faecium (1 hit).

NECTIN4 (PVRL4) is a cell adhesion molecule member of the nectin and nectin-like protein family. NECTIN4 is involved in the formation and maintenance of adherens junctions and is expressed in various tissues, including the lungs, kidneys and salivary glands. The overexpression of NECTIN4 has been shown in several types of cancer, including breast, lung, ovarian and pancreatic cancers. It is associated with tumor progression and poor patient prognosis. Given its involvement in various aspects of cancer progression and metastasis, NECTIN4 has emerged as a potential therapeutic target for cancer treatment (Delpeut et al., 2014).

**FOLH1 and TRAP1 (a mitochondria homologue of HSP90)**

The FOLH1 epitope “KAFLDEL” was found to be highly similar to TRAP1 epitope “KAFLDAL” (Fig. 2 and Fig. 3E) with experimental evidence of presentation at the MHC Class-I locus with supporting *quantitative binding* experimental data to HLA-A02 (IEDB epitope 1635984). The TRAP1-derived epitope “KAFLDAL” was presented at MHC class-II (IEDB epitope 1299293) and was highly homologous with 217 human pathogen epitopes, including Salmonella enterica (174 hits), Klebsiella pneumoniae (18 hits), Escherichia coli (13 hits), Citrobacter freundii (1 hit), Enterobacter ludwigii (1 hit), Enterobacter asburiae (1 hit), Treponema pallidum (4 hits), Mycobacterium tuberculosis (4 hits) and Treponema pallidum (1 hit).

Glutamate carboxypeptidase-II encoded by Folate hydrolase 1 (FOLH1), also known as prostate-specific membrane antigen (PSMA), is a type-II transmembrane protein that is highly expressed in prostate epithelial cells and is overexpressed in various solid tumors including prostate, bladder, kidney and lung cancers. It has been implicated in cancer cell proliferation, migration, invasion and angiogenesis. FOLH1/PSMA-targeted agents, including small molecule inhibitors, antibodies and radiolabeled ligands have shown promising results in preclinical and clinical studies for the diagnosis and treatment of various cancers (L. Ma et al., 2022).

Tumor necrosis factor Receptor-Associated Protein 1 (TRAP1) is a mitochondrial chaperone homologue of HSP90 which plays an important role in regulating cellular metabolism, stress response and apoptosis. TRAP1 is overexpressed in many types of cancer cells and may contribute to cancer progression and drug resistance and is an emerging target in cancer (Park et al., 2020; Vartholomaiou et al., 2017; Wengert et al., 2022).

**GSN and HSP90 co-chaperone CDC37L**

The GSN epitope “SEAEKTGAQEL” was identified to be highly similar to the HSP90 co-chaperone CDC37L-derived epitope “LEAEKKGAL” (Fig. 3F). Experimental evidence of presentation at HLA Class I and II was identified for the GSN-derived epitope in 11 IEDB records. The corresponding CDC37L-derived epitope was shown to be presented at MCH class-II in two IEDB records and had 22 homologs in human pathogens, including Staphylococcus pseudintermed (4 hits), Flavobacterium psychrophilum (2 hits), Salmonella enterica (1 hit) and Bacillus cereus group (11 hits).

Gelsolin (GSN) is an actin-binding protein which regulates cell motility and morphology, as well as apoptosis and inflammation. It is dysregulated in various types of cancer, including breast, lung, prostate and colorectal cancer, among others, and is involved in tumor progression and metastasis by promoting the invasion and migration of cancer cells through the regulation of the actin cytoskeleton (X. Ma et al., 2016).

## DISCUSSION

Autoantibodies and self-reactive T cells exist in normal individuals. Yet, most individuals do not develop autoimmunity nor are they protected against cancer-associated autoantigens. Arguably, self-reactivity is not sufficient for the molecular mimicry to become effective. The orchestrated immune response in febrile infection, including pro-inflammatory processes such as interferon (IFN) signaling and tumor necrosis factor (TNF) signaling is shared in cancer (Karin & Greten, 2005). This response may also radically reprogram the peptide presentation repertoire and reprogram peptide immunogenicity during inflammation (Faridi et al., 2020). Furthermore, the transient upregulation of HSPs in inflammation, e.g., HSP90 (Iheagwara et al., 2014), may act as an adjuvant danger signal and improve antigen presentation (Wahl et al., 2010) to further amplify these pro-inflammatory phenotypes (van Wijk & Prakken, 2010). The synergic accumulation of these effects could ultimately push the cross-reactive T or B cells over the self-tolerance barriers (Pockley et al., 2008; W Van Eden et al., 2007) and consequently confer cancer immunity. Therefore, in this context, it is plausible to study HSPs as mediators for the emergence of effective molecular mimicry.

MUC1 is temporarily upregulated in primary mammary epithelial cells in response to tissue inflammation (Jacqueline, Lee, et al., 2020) and increased anti-MUC1 antibody titers in women are associated with a lower risk of developing ovarian cancer (Cramer et al., 2005). This gene is a candidate for cancer vaccines in numerous clinical trials (Deng et al., 2013), including ovarian cancer (Gray et al., 2016), colon (Finn, 2021; Kimura et al., 2013), recurrent colorectal adenoma (Schoen et al., 2022), lung cancer (Finn et al., 2023) and early-stage breast cancer (Apostolopoulos et al., 2006). Increased anti-MUC1 titer in individuals with multiple episodes of mumps infections in women is associated with a lower risk of ovarian cancer (Cramer et al., 2010b; Finn, 2014; Menczer et al., 1979; Pinheiro et al., 2010) and is correlated with higher overall survival in breast cancer patients (Fremd et al., 2016). The transient overexpression of self-antigens, such as MUC1, in inflamed infected tissues may be another important factor which facilitates the effective implementation of molecular mimicry. This was experimentally demonstrated in murine models of acute lymphocytic choriomeningitis virus (LCMV) infection (Jacqueline et al., 2022) in which induced antibodies and T cells cross-reacted with tumors and conferred immunity to further tumor challenges. This further supports the core idea examined in this study that suggests cancer-associated antigens may be carrying epitope candidates involved in molecular mimicry.

The overexpression of HSPs mediates two hallmarks of cancer, i.e., cell proliferation and senescence, and is implicated in tumor cell proliferation, differentiation, invasion and metastasis (Kumar et al., 2020; Seclì et al., 2021). Extracellular HSPs have been found in the exosomes, membrane surfaces and sera of patients with various pathological conditions and are biomarkers for cancer progression, immune surveillance, immune evasion, metastasis and cancer treatment (Albakova et al., 2021; Caruso Bavisotto et al., 2022; Eguchi et al., 2020; Krawczyk et al., 2020; Poggio et al., 2021). As such, targeting HSPs as cancer-associated antigens has received attention in recent years (Albakova et al., 2022; Cyran & Zhitkovich, 2022). In addition to HSP inhibitors, neutralizing antibodies have shown effectiveness in targeting HSPs (Sidera et al., 2011; Stellas et al., 2007, 2010; Stivarou et al., 2016) and patient-derived antibodies cross-reactivity with HSPs (Devarakonda et al., 2015; Trieb et al., 2000). Our results may extend the scope of neutralizing HSPs to antibodies that target both HSPs and CAAs at their shared epitopes.

Our method could be expanded to other cancer-associated antigens, cancer-specific antigens and cancer testis antigens. In addition to HSPs, it is also possible other proteins and cytokines are upregulated in infection, recognized by the immune system and share epitopes with CAAs. In this context, it is conceivable many TCRs and BCRs in infectious diseases and cancer databases have sequence similarities, as we recently observed in mesothelioma cancer (unpublished).

Our method could also be used for shedding light on underlying mechanisms governing the success or failure of therapies which induce hyperthermia or fever, e.g., by employing external stimuli such as mistletoe extracts for cancer treatment (Køstner et al., 2015; Reuter et al., 2018). It can also be used to better understand the effect of infection in enhancing immunotherapies. An emerging body of research suggests pre-existing immunity to pathogen antigens is crucial in response to immunotherapy (Conti, 2021). For instance, melanoma patients undergoing anti-PD1 treatment who had a high IgG titer for cytomegalovirus (CMV) show prolonged progression-free survival (Chiaro et al., 2021). Further, inducing infection at the tumor site, e.g., by intratumoral injection of the seasonal flu shot, has been shown to reprogram immunologically cold tumors into hot, thus improving the efficacy of immunotherapies (Newman et al., 2020). A similar approach to the one proposed here may be used to identify the role of molecular mimicry in these scenarios.

Molecular mimicry of HSP-derived epitopes may also be useful in the context of autoimmunity. HSP90 has been implicated in the pathogenesis of autoimmunity and is associated with altered idiotypic regulation of the anti-HSP90 IgG autoantibodies in patients with autoimmune disorders like systemic lupus erythematosus (SLE) (Kenderov et al., 2002). Surface expression of HSP90 on peripheral blood mononuclear cells was found in approximately 25% of the SLE patients with active disease (Kenderov et al., 2002). These autoantibodies are also biomarkers of autoimmune retinopathy (Adamus et al., 2013). Moreover, normal human IgG has been found to contain considerable amounts of low-affinity anti-HSP90 natural auto-antibodies (Kenderov et al., 2002) and the expression of extracellular HSP90 is a molecular signature of T cell activation and a biomarker for T cell activation in autoimmune disease (Scarneo et al., 2022). Application of methods proposed in this study may reveal new insights for pathogenesis of these diseases.

Premalignant cells are constantly generated due to numerous environmental carcinogens and random mutations. In a so-called equilibrium state (Schreiber, 2011), our immune system eradicates the majority of these cells. However, when malignant cells escape this equilibrium and establish the tumor, their elimination becomes a challenge for the immune system. The established immune suppressive microenvironment, tumor genome instability, and downregulation of antigen presentation machinery are among the main reasons that give a near “infinite ability” for tumor to evolve and evade therapies (Abbosh et al., 2023; Cells, 2010). Therefore, it is an increasingly accepted that prevention, e.g. in high-risk individuals, is favorable to cancer treatment (Çuburu et al., 2022; Finn, 2018; Lollini et al., 2006). By approaching infection as a surrogate, our approach presents a step towards this goal and provides a mechanistic understanding of long-term immunity to cancer (Jacqueline et al., 2018).

The idea of using HSPs as surrogates between infection and cancer based on molecular mimicry has several immediate applications. First, the epitopes shared between HSPs and CAAs can be used as candidates for cancer vaccines. Furthermore, the identified epitopes may be used to identify corresponding autoreactive T cell and B cell specificities that, in turn, are used as therapeutic candidates for TCR-T cell therapies and antibody drugs, respectively. While the experimental evidence obtained from the IEDB database and homolog epitopes from human pathogens supports potential immunogenicity of identified epitopes, a lack of *in-vitro* and *in-vivo* experimental validation renders the results of this study as only plausible hypotheses. The validation of these epitopes is the subject of our ongoing work.

## METHODS AND MATERIALS

***HSP selection:*** HSPs are named according to their molecular weight, which ranges from 17kDa small HSP families to over 100kDa. They are classified into families: the HSP100, HSP90, HSP70, HSP60, and HSP40. We searched UniProt (Coudert et al., 2023) and selected 281 heat-shock proteins and fragments as enumerated in Supplementary Table 1A.

***CAA selection:*** A list of 178 tumor-associated antigens was prepared, prioritized by the National Cancer Institute (NCI) (Cheever et al., 2009). Additionally, the cancer-associated antigens (overexpressed) identified in Cancer Antigen Peptide Database (https://caped.icp.ucl.ac.be/) were included in this study (Supplementary Table 1B).

***IEDB annotation:*** The IEDB epitope version 3 database, released in January 2023 (https://www.iedb.org), was downloaded and used for annotation of epitopes.

***Human pathogen annotation*:** A list of 46 human pathogen species was prepared (Supplementary Table 1C) according to Human Pathogens released by the NCBI Pathogen Detection database (www.ncbi.nlm.nih.gov/pathogens, January 2023).

***p-value computation:*** The p-value for stretches of 6–9 identical aligned amino acids (66.6 to 100% homology epitopes between microorganisms and human gene was previously reported in extremely low ($<1.56 \times 10^{-8}$) (Tagliamonte et al., 2023). Since the method by which this very low p-value was calculated could not be confirmed, a computational simulation was performed, involving a search for 100,000 epitopes of length seven or more shared between random pairs of human proteins. Approximately 72% of the time, shared epitopes were identified with a maximum of 1 residue difference. Then, these epitopes were mapped to the IEDB epitope database; any pair which had at least one record in IEDB with matched genes for both epitopes was retained. Four in 100,000 epitopes passed the criteria (p-value ~ $10^{-5}$).

## ACKNOWLEDGMENT

The author would like to thank Yoav Litvin Ph.D. (https://yoavlitvin.com) for editing this text and for invaluable comments that improved the manuscript.

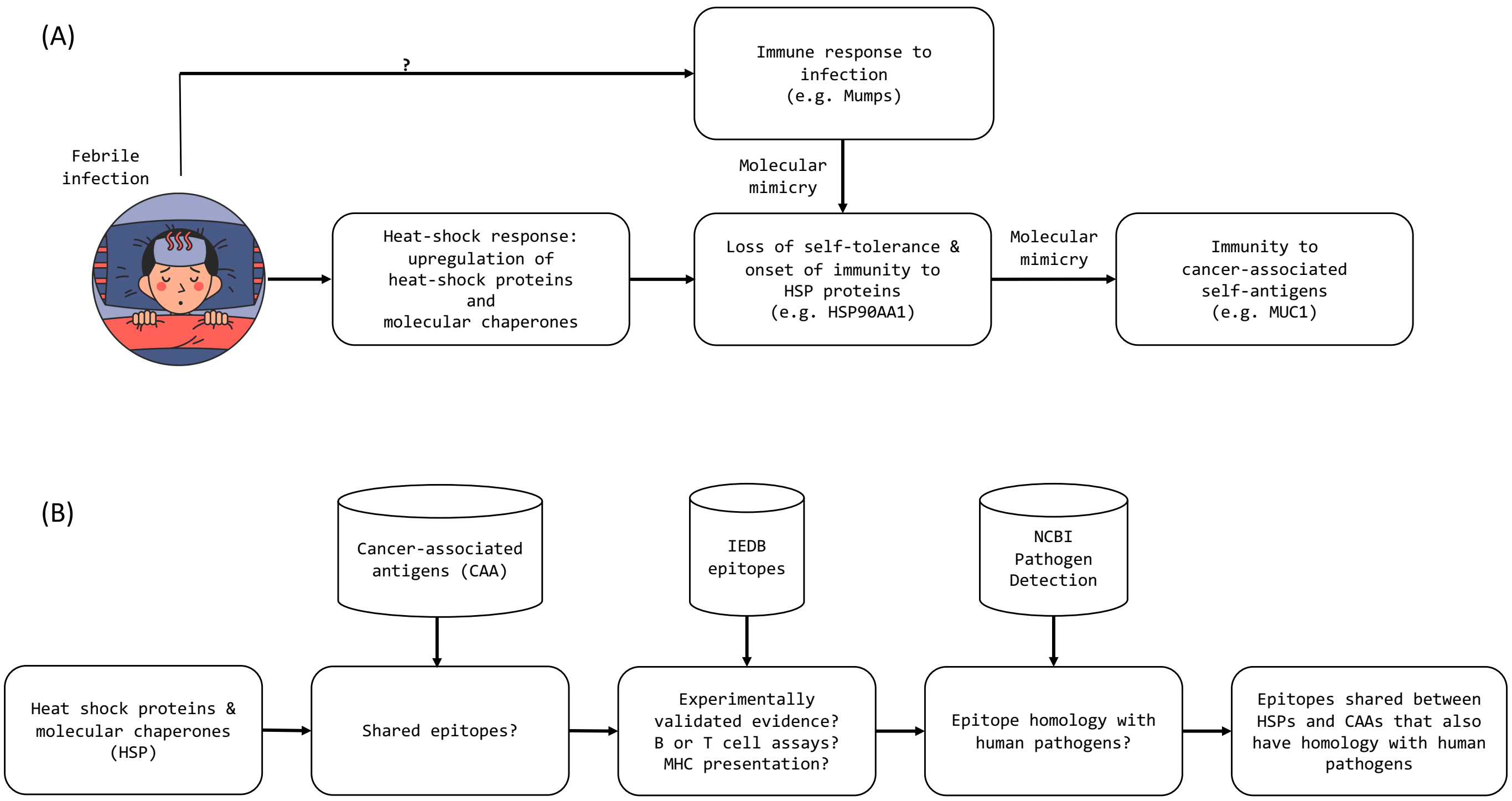

Fig. 1 (A) The hypothesis examined in present study that suggest heat-shock proteins mediate cancer immunity during febrile infection through molecular mimicry (B) The method used to examine the hypothesis.

| Heat-shock protein (HSP) | Cancer-associated antigen (CAA) | Epitope pairs | IEDB record for HSP epitopes | | | | HSP epitope hits | HSP Molecular weight |
|---|---|---|---|---|---|---|---|---|
| | | Total: 94 | MHC Class-I | MHC Class-II | T-cell assay | B-cell assay | Human pathogen | |
| HSPE1 | PTPRS, MYO1B | 2 | 30 | 46 | | | 46 | 10kDa |
| HSPB1, HSPB2, HSPB6, HSPB7, HSPB8 | OGT,EPHA2,KLK3,EEF2, MUC3A,MUC5AC | 6 | 24 | 26 | | | 29 | 28kDa |
| HSPD1 | CALCA,KIF20A,EPHA3 | 3 | 49 | 103 | 8 | 6 | 258 | 60kDa |
| HSPA14,DNAJA1,DNAJA2,DNAJA3,DNAJA4,DNAJB11,DNAJB6,DNAJC2,DNAJC3, DNAJC7,HSPA7,HSPA12A,HSPA12B, HSPA13,HSPA14,HSPA4,HSPA4L,HSPA6 | CFLAR,PTPRK,GSN,PTPRC,FN1,GSN,CDC27,FN1,EWSR1,SIRT2,PTPRA,GSN, PSAT1,EZH2,LGSN,CDK12,ERBB2,CDK4,CALCB,COL20A1,CALCA,CDK4, TYR,PTPRG, UBR4,ERBB2,PTPRA,FN1,CTNNB1,BRAF,LCK,EPHA3,IL13RA2, PTPRS,ERBB2,PTPRS,PTPRS,PTPRS,KDR | 37 | 172 | 287 | | | 1360 | 70kDa |
| HSPA1L,HSPA1B,HSPA1A, HSPA6,HSPA8,HSPA2 | PDGFRB,FN1,MMP2, FNDC3B,CCDC110 | 5 | 90 | 195 | 1 | 5 | 2187 | 71kDa |
| TRAP1 | MYO1B,FOLH1,CYP1B1, ENAH,FN1 | 5 | 11 | 20 | | | 221 | 75kDa |
| CDC37,CDC37L1,AHSA1 | OGT,MUC5AC,EEF2,GSN,GPNMB,MAGEC1,CDC27,GPNMB | 8 | 22 | 22 | | | 46 | HSP90-associated |
| HSP90B1,HSP90AA1,HSP90AB1, HSP90AB1,HSP90AB3P,HSP90AA4P | MMP2,TEK,ACTN4,ACTN1,EFTUD2, MUC1,NECTIN4,GPC3,UBR4 | 9 | 81 | 138 | | | 1057 | 90kDa |
| HSPH1,HSPA4 | ERBB2,GAS7,PTPRD,PTPRS,CSPG4,SYK,EZH2, GSN,EEF2,PTPRG,EFTUD2 | 10 | 87 | 124 | | | 367 | 105KdA |
| HSPA2,HSF2BP,HSF1,HSF2 | MMP2,MGAT5,UBR4,MDM2, GSN,RHOC,MDM2 | 9 | 15 | 21 | | | 145 | HSF |

Table-1: A summary of epitopes identified on HSP 90kDa proteins that are shared with cancer-associated antigens. The IEDB record and human pathogen hit counts are only for HSP proteins (corresponding numbers for CAAs are available in Supplementary Table 3).

```
                                                                                                                      (FOLH1) KAFLDEL
HSP90AA1   MPEETQTQDQPMEEEEVETFAFQAEIAQLMSLIINTFYSNKEIFLRELISNSSDALDKIRYESLTDPSKLDSGKELHINLIPNKQDRTLTIVDTGIGMTKADLINNLGTIAKSGTKAFMEAL
HSP90AB1   MPEEVHHG-----EEEVETFAFQAEIAQLMSLIINTFYSNKEIFLRELISNASDALDKIRYESLTDPSKLDSGKELKIDIIPNPQERTLTLVDTGIGMTKADLINNLGTIAKSGTKAFMEAL
                                                                                                                      (TRAP1*) KAFLDAL
                           YSLFLVA (MMP2)
HSP90AA1   QAGADISMIGQFGVGFYSAYLVAEKVTVITKHNDDEQYAWESSAGGSFTVRTDTGEPMGRGTKVILHLKEDQTEYLEERRIKEIVKKHSQFIGYPITLFVEKERDKEVSDDEAEEKEDKEEE
HSP90AB1   QAGADISMIGQFGVGFYSAYLVAEKVVVITKHNDDEQYAWESSAGGSFTVRADHGEPIGRGTKVILHLKEDQTEYLEERRVKEVVKKHSQFIGYPITLYLEKEREKEISDDEAEEEKG---E
                           YSAFLVA (HSP90B1*)
                                                                                                                        NRKVKNN (TEK)
HSP90AA1   KEKEEKESEDKPEIEDVGSDEEEEKKDGDKKKKKKIKEKYIDQEELNKTKPIWTRNPDDITNEEYGEFYKSLTNDWEDHLAVKHFSVEGQLEFRALLFVPRRAPFDLFENRKKKNNIKLYVR
HSP90AB1   KEEEDKDDEEKPKIEDVGSDEEDDSGKDKKKKTKKIKEKYIDQEELNKTKPIWTRNPDDITQEEYGEFYKSLTNDWEDHLAVKHFSVEGQLEFRALLFIPRRAPFDLFENKKKKNNIKLYVR

                                                     (UBR4) IVQKCLE    LAEDIEN (EFTUD2)
HSP90AA1   RVFIMDNCEELIPEYLNFIRGVVDSEDLPLNISREMLQQSKILKVIRKNLVKKCLELFTELAEDKENYKKFYEQFSKNIKLGIHEDSQNRKKLSELLRYYTSASGDEMVSLKDYCTRMKENQ
HSP90AB1   RVFIMDSCDELIPEYLNFIRGVVDSEDLPLNISREMLQQSKILKVIRKNIVKKCLELFSELAEDKENYKKFYEAFSKNLKLGIHEDSTNRRRLSELLRYHTSQSGDEMTSLSEYVSRMKETQ

                                                                                                      DKKVLKV (GPC3)
HSP90AA1   KHIYYITGETKDQVANSAFVERLRKHGLEVIYMIEPIDEYCVQQLKEFEGKTLVSVTKEGLELPEDEEEKKKQEEKKTKFENLCKIMKDILEKKVEKVVVSNRLVTSPCCIVTSTYGWTANM
HSP90AB1   KSIYYITGESKEQVANSAFVERVRKRGFEVVYMTEPIDEYCVQQLKEFDGKSLVSVTKEGLELPEDEEEKKKMEESKAKFENLCKLMKEILDKKVEKVTISNRLVSSPCCIVTSTYGWTANM

                                                   (ACTIN1,ACTIN4) YETATLS   SLEDPST (MUC1)                  SAAVTSE (NECTIN4)
HSP90AA1   ERIMKAQALRDNSTMGYMAAKKHLEINPDHSIIETLRQKAEADKNDKSVKDLVILLYETALLSSGFSLEDPQTHANRIYRMIKLGLGIDEDDPTADDTSAAVTEEMPPLEGDDDTSRMEEVD
HSP90AB1   ERIMKAQALRDNSTMGYMMAKKHLEINPDHPIVETLRQKAEADKNDKAVKDLVVLLFETALLSSGFSLEDPQTHSNRIYRMIKLGLGIDEDEVAAEEPNAAVPDEIPPLEGDEDASRMEEVD

(*) these epitopes are from other HSP families.
```

Fig. 2. An alignment for two HSP90 protein family members. Identified epitopes are in Red. Tumor associated antigens in Red appear in parentheses. The epitopes besides genes in Green parentheses are from other members of HSP family. Residues in shared epitopes colored in Black are evolutionarily fully conserved (according to MAFFT alignment).

| | | | |
|---|---|---|---|
| epitope 2071116 | ----------------SLEDPSTDYYQELQR----- | Class II | CAA IEDB epitope |
| MUC1 | SFFFLSFHISNLQFNSSLEDPSTDYYQELQRDISEM | | Epitope pair |
| | *****.* | | |
| HSP90AB1 | DLVVLLFETALLSSGFSLEDPQTHSNRIYRMIKLGL | | |
| epitope 1278711 | ----------------SLEDPQTHANRIYRM----- | Class II | HSP IEDB epitopes |
| epitope 1255422 | -----------LSSGFSLEDPQTHSNR--------- | Class II | |
| epitope 1255419 | -----------LSSGFSLEDPQTHANRIYRM----- | Class II | |
| epitope 1278713 | ----------------SLEDPQTHSNRI-------- | Class II | |
| epitope 1828420 | ------------SSGFSLEDPQTHSN---------- | Class II | |
| epitope 1828419 | ------------SSGFSLEDPQTHAN---------- | Class II | |
| epitope 1255420 | -----------LSSGFSLEDPQTHANRIYRMI---- | Class II | |
| epitope 1116982 | ----------------SLEDPQTHAN---------- | Class II | |
| epitope 1159331 | -----------LSSGFSLEDPQTHSN---------- | Class II | |
| epitope 1211554 | ------FETALLSSGFSLEDPQTHSNRIY------- | Class II | |
| epitope 1828421 | ------------SSGFSLEDPQTHSNRIY------- | Class II | |
| epitope 1278712 | ----------------SLEDPQTHANRIYRMIKLGL | Class II | |
| epitope 1278710 | ----------------SLEDPQTHANRIY------- | Class II | |
| epitope 1109250 | -----------LSSGFSLEDPQTH------------ | Class II | |
| epitope 1255418 | -----------LSSGFSLEDPQTHANRIYR------ | Class II | |
| epitope 1278714 | ----------------SLEDPQTHSNRIYRM----- | Class II | |
| epitope 1392610 | ---------ALLSSGFSLEDPQTHANRIY------- | Class II | |
| epitope 1255421 | -----------LSSGFSLEDPQTHS----------- | Class II | |
| epitope 1255417 | -----------LSSGFSLEDPQTHANRIY------- | Class II | |
| epitope 1159330 | -----------LSSGFSLEDPQTHAN---------- | Class II | |
| epitope 1255425 | -----------LSSGFSLEDPQTHSNRIYRMI---- | Class II | |
| epitope 1392611 | ---------ALLSSGFSLEDPQTHSNRIY------- | Class II | |
| epitope 1255424 | -----------LSSGFSLEDPQTHSNRIYRM----- | Class II | |
| epitope 1116983 | ----------------SLEDPQTHSN---------- | Class II | |
| epitope 1278715 | ----------------SLEDPQTHSNRIYRMIKLGL | Class II | |
| epitope 634103 | ----------LLSSGFSLEDPQTHSNR--------- | Class I | |
| epitope 941526 | ----------------SLEDPQTHSNRIYR------ | Class I | |
| epitope 1255416 | -----------LSSGFSLEDPQT------------- | Class I | |
| epitope 865370 | ----------------SLEDPQTH------------ | Class I | |
| epitope 1048911 | -------------SGFSLEDPQTHSNR--------- | Class I | |
| epitope 541532 | DLVILLYETALLSSGFSLEDPQTHANR--------- | Class I | |
| epitope 1348888 | ---------ALLSSGFSLEDPQTHS----------- | Class I | |
| epitope 1116984 | ----------------SLEDPQTHSNR--------- | Class I | |
| epitope 629229 | ---------ALLSSGFSLEDPQTHANR--------- | Class I | |
| epitope 1348889 | ---------ALLSSGFSLEDPQTHSN---------- | Class I | |
| epitope 541534 | DLVVLLFETALLSSGFSLEDPQTHSNR--------- | Class I | |
| epitope 427324 | ----------------SLEDPQTHSNRIY------- | Class I | |
| epitope 629230 | ---------ALLSSGFSLEDPQTHSNR--------- | Class I | |
| epitope 925964 | ---------------FSLEDPQTH------------ | Class I | |
| Salmonella enterica | DLVVLLFATALLSSGFSLEDPQTHSNRIYRMIKLGL | 785 hits | Human pathogen |
| Streptococcus agalactiae | DLVILLYETALLSSGFSLEDPQTHANRIYRMIKLGL | 19 | |
| Burkholderia cepacia complex | ----------------SLEDPQT------------- | 2 | |
| Escherichia coli | DLVVLLFETALLSSGFSLDDPQTHSNRIYRMIKLGL | 19 | |
| Klebsiella pneumoniae | DLVVLLFETALLSSGFSLDDPQTHSNRIYRMIKLGL | 165 | |
| Enterobacter hormaechei | DLVILLFETALLSSGFSLDDPQTHSNRIYRMI---- | 39 | |
| Enterobacter cloacae | ----------------SLEDPETH------------ | 4 | |
| Pseudomonas putida | ---------------FSLEKPQTH------------ | 2 | |
| Enterococcus faecium | DLVILLFETSLLSSGFTLDDPQTHSNRIYRMIKLGL | 25 | |
| Stenotrophomonas maltophilia | ---------------FSLENPQT------------- | 5 | |
| Acinetobacter baumannii | -------------SGFSLEDGRTH------------ | 1 | |

Fig. 3A: HSP90-epitope is shared with epitope from MUC1 and has several experimentally validated IEDB records that show it is presented at MHC class-I and II. It is also homolog with several human pathogens in NCBI Pathogen Detection database.

```
epitope 1305673              YPEDLFIDKKVLKVAH------       Class II   } CAA IEDB epitope

GPC3                         YPEDLFIDKKVLKVAHVEHEET                  }
                                    **** **                          } Epitope pair
HSP90AB1                     KLMKEILDKKVEKVTISNRLVS                  }

epitope 1885807              ----EILDKKVEKVTISNRL--       Class II   }
epitope 1903229              ---KEILDKKVEKVTISNRL--       Class II   }
epitope 1963528              -------DKKVEKVTISNRLVS       Class II   }
epitope 1967633              ------LDKKVEKVTISNRLV-       Class II   }
epitope 1247440              ------LDKKVEKVTISNRL--       Class II   }
epitope 27000                -----ILDKKVEKV--------       Class I *  } HSP IEDB epitopes
epitope 1319425              ---KEILDKKVEKV--------       Class I    }
epitope 1567453              ----EILDKKVEKVT-------       Class I    }
epitope 456002               ------LDKKVEKV--------       Class I    }
epitope 541507               -------DKKVEKV--------       Class I    }
epitope 758788               -------DKKVEKVTISNRL--       Class I    }

Clostridioides difficile     ---EIINDKKVEKV--------       231 hits   }
Salmonella enterica          ---KEILDKKVEKVTISNRLVS       12         }
Enterobacter hormaechei      ---KEILDKKVEKVTVSNRLVS       6          }
Vibrio vulnificus            -------DKKVEKV--------       1          }
Acinetobacter baumannii      -------DKKVEKV--------       1          }
Legionella pneumophila       ----EVLDKKTEKVTISNRL--       5          }
Clostridium botulinum        ----EILDKKEEKVTIT-----       6          } Human pathogen
Escherichia coli             ------LEKKVEKV--------       2          }
Enterobacter kobei           -----ILNKKVETVT-------       2          }
Pseudomonas putida           ------LDKKIEKV--------       1          }
Streptococcus pneumoniae     ----DILEKKVEKV--------       52         }
Streptococcus agalactiae     -----ILEKKVEKVVVSNRLV-       4          }
Flavobacterium psychrophilum -----ILDKKVEHV--------     12         }
Bacillus cereus group        -----ILNKQVEKVT-------       1          }

(*): MHC qualitative binding
```

Fig. 3B: HSP90 epitope shared with epitope from GPC3. IEDB epitope 27000 presents evidence of MHC Class-I quantitative binding for the shared epitope.

```
epitope 1227192              -----------GYSLFLVAAHEFGHA-   Class II  } CAA
epitope 1759811              -----------GYSLFLVAAHEFG---   Class II  } IEDB epitope

MMP2                         DDRKWGFCPDQGYSLFLVAAHEFGHAM             }
                                         ** ****  :                  } Epitope pair
HSP90B1                      TSELIGQFGVGFYSAFLVADKVIVTSK             }

epitope 642641               --------GVGFYSAFLVADK------   Class II  } HSP IEDB epitopes

Klebsiella pneumoniae        ---LIGQFGVGFYSAFLVADKVTV---   1 hit     }
Flavobacterium psychrophilum DVSLIGQFGVGFYSAFLVADAVTV---   1         }
Enterococcus faecium         ------------YSAFLVA--------   1         }
Clostridium botulinum        -------FGVGFYSAFLVADK------   2         }
Serratia marcescens          ---LIGQFGVGFYSAFIVADKVIV---   1         } Human pathogen
Escherichia coli             DSQLIGQFGVGFYSAFIVADKVIVRTR   2         }
Pseudomonas aeruginosa       DASLIGQFGVGFYSAFIVADKVEVFTR   4         }
Shigella                     DSQLIGQFGVGFYSAFIVADKVIVRTR   2         }
Salmonella enterica          DSQLIGQFGVGFYSAFIVADKVIVRTR   6         }
```

Fig. 3C: HSP90 epitope shared with epitope from MMP2

| | | | |
|---|---|---|---|
| epitope 1282685 | ------SSRSFKHSR**SAAVTSE**----------------- | Class II | CAA IEDB epitope |
| NECTIN4 | TEVKGTTSSRSFKHSR**SAAVTSE**FHLVPSRSMNGQPLTCV | | Epitope pair |
| | *****.* | | |
| HSP90AA1 | LGLGIDEDDPTADDT**SAAVTEE**MPPLEGDDDTSRMEEVD | | |
| epitope 1115594 | ---------------**SAAVTEE**MPPLEGDDDTSRM---- | Class II | HSP IEDB epitopes |
| epitope 1115596 | ---------------**SAAVTEE**MPPLEGDDDTSRMEEVD | Class II | |
| epitope 1163516 | ---------PTADDT**SAAVTEE**MPPLEGDDDTSRMEEVD | Class II | |
| epitope 137823 | --------DPTADDT**SAAVTEE**MPP------------- | Class II | |
| epitope 1598086 | -----------ADDT**SAAVTEE**MPP------------- | Class II | |
| epitope 543665 | LGLGIDEDDPTADDT**SAAVTEE**MPPLEGDDDTSR----- | Class I | |
| epitope 641694 | --------DPTADDT**SAAVTEE**MPPLEGDDDTSRMEEVD | Class I | |
| Pseudomonas aeruginosa | ---------------**SAAVTEE**----------------- | 36 hits | Human pathogen |
| Bacillus cereus group | ---------------**SAAVTEE**----------------- | 19 | |
| Streptococcus agalactiae | ---------------TP**AVTEE**MPPLEGDDDTSRMEEVD | 1 | |
| Streptococcus agalactiae | ---------------TP**AVTEE**MPPLEGDDDTSRM---- | 1 | |
| Streptococcus agalactiae | -----------AEEATP**AVTEE**MPPLEGDDDTSRMEEVD | 1 | |
| Enterococcus faecium | --------------T**SA**PS**T**D**E**IPPLEGDDDASRMEEVD | 1 | |

Fig. 3D: HSP90 epitope shared with epitope from NECTIN4 (PVRL4)

| | | | |
|---|---|---|---|
| epitope 1635984 | ----------NM**KAFLDEL**-- | Class I | CAA IEDB epitope |
| FOLH1 | NEATNITPKHNM**KAFLDEL**KA | | Epitope pair |
| | *****.* | | |
| TRAP1 | VSNLGTIARSGS**KAFLDAL**QN | | |
| epitope 1299293 | VSNLGTIARSGS**KAFLDAL**QN | Class II | HSP IEDB epitope |
| Salmonella enterica | ------------**KAFLDAL**-- | 174 hits | Human pathogen |
| Klebsiella pneumoniae | ------------**KAFLDAL**-- | 18 | |
| Escherichia coli | ------------**KAFLDA**L-- | 13 | |
| Citrobacter freundii | ------------**KAFLDAL**-- | 1 | |
| Enterobacter ludwigii | ------------**KAFLDAL**-- | 1 | |
| Enterobacter asburiae | ------------**KAFLDAL**-- | 1 | |
| Treponema pallidum | --NLGTIARSGT**KAFL**STL-- | 4 | |
| Mycobacterium tuberculosis | --NLGTIARSGS**K**E**FL**E**AL**-- | 4 | |
| Treponema pallidum | --NLGTIARSGT**KAFL**EQ**L**-- | 1 | |

Fig. 3E: Epitope from HSP90 mitochondrial homolog TRAP1 shared with epitope from cancer-associated antigen FOLH1

```
epitope 439262                  -LWVGTGASEAEKTGAQEL------------         Class I
epitope 441683                  --WVGTGASEAEKTGAQEL------------         Class I
epitope 1124438                 YLWVGTGASEAEKTGAQEL------------         Class II
epitope 1201751                 ---------EAEKTGAQELLRVLRAQPVQ--         Class II
epitope 1201752                 ---------EAEKTGAQELLRVLRAQPVQVA         Class II      CAA
epitope 1217592                 ------GASEAEKTGAQELLRVLRAQPVQ--         Class II      IEDB epitope
epitope 1217593                 ------GASEAEKTGAQELLRVLRAQPVQVA         Class II
epitope 1275300                 --------SEAEKTGAQELLRVLRAQPV---         Class II
epitope 1286298                 -----TGASEAEKTGAQELLRVLRAQPVQ--         Class II
epitope 1294626                 ---VGTGASEAEKTGAQEL------------         Class II
epitope 1302597                 --WVGTGASEAEKTGAQELL-----------         Class II

GSN                             YLWVGTGASEAEKTGAQELLRVLRAQPVQVA
                                         ****.**  *                           epitope pair
Hsp90 co-chaperone CDC37L1       YLILWCFHLEAEKKGA--LMEQIAHQAVVMQ

epitope 1247858                 --------LEAEKKGA--LMEQIAH------         Class II      HSP
epitope 1247859                 --------LEAEKKGA--LMEQIAHQ-----         Class II      IEDB epitopes

Staphylococcus pseudintermed    --------LEAEKDGA--LAME---------         4 hits
Flavobacterium psychrophilum    ---------EAEKKGA---------------         2             Human
Salmonella enterica             --WL---AEEAEKKGA--L------------         1             pathogen
Bacillus cereus group           ---------EAEKKGA---------------         11
```

Fig. 3F: HSP90 co-chaperone CDC37L1 epitope shared with GSN-derived epitope